# Origin of the dense core mass function in contracting filaments

**Philip C. Myers**

Harvard-Smithsonian Center for Astrophysics, 60 Garden Street,

Cambridge MA 02138 USA

pmyers@cfa.harvard.edu

**Abstract.** Mass functions of starless dense cores (CMFs) may arise from contraction and dispersal of core-forming filaments. In an illustrative model, a filament contracts radially by self-gravity, increasing the mass of its cores. During this contraction, FUV photoevaporation and ablation by shocks and winds disperse filament gas and limit core growth. The stopping times of core growth are described by a waiting-time distribution. The initial filament column density profile and the resulting CMF each match recent *Herschel* observations in detail. Then low-mass cores have short growth ages and arise from the innermost filament gas, while massive cores have long growth ages and draw from more extended filament gas. The model fits the initial density profile and CMF best for mean core density $2 \times 10^4$ cm$^{-3}$ and filament dispersal time scale 0.5 Myr. Then the typical core mass, radius, mean column density, and contraction speed are respectively 0.8 solar masses, 0.06 pc, $6 \times 10^{21}$ cm$^{-2}$, and 0.07 km s$^{-1}$, also in accord with observed values.

1. **Introduction**

Dense cores are believed to be birth sites of stars, because they are frequently associated with protostars (Beichman et al. 1986, Enoch et al. 2006, Jørgensen et al. 2008). The relation of core and protostar masses is less clear, however. On one hand, the mass distribution of observed starless dense cores (CMF), resembles the initial mass function of stars (IMF), in its turnover at low mass and in its power-law slope at high mass (Motte et al. 1998, Alves et al. 2007, Könyves et al. 2010). These properties indicate that the IMF can be approximated by multiplying each core mass in a CMF by a constant factor $\varepsilon < 1$.

This similarity of the CMF to the IMF may indicate that the mass of a starless core determines the mass of the protostar it will form (Alves et al. 2007, André et al. 2010). In this interpretation, the IMF is set at an early stage of cloud evolution by processes which set the CMF.

On the other hand, observed core mass does not have a unique definition. Its value depends on resolution and on sensitivity to density and column density (Swift & Williams 2008), and on how the contribution of "background" gas to core mass is counted. Also, cores in clusters blend in projection (Kainulainen et al. 2011, Michel et al. 2011), and cores of differing mass may produce protostars at different rates (Hatchell & Fuller 2008). Some observed cores harbor multiple stars, while others disperse before forming any stars (Myers 2010). These points suggest that core and protostar masses may have a statistical relationship which is neither deterministic nor coincidental.

Although the exact relation between the CMF and the IMF remains unclear, it is important for our understanding of star-forming clouds to understand the origin of the CMF. Many authors have accounted for the CMF with models of gravoturbulent fragmentation (Klessen & Burkert 2001, Padoan & Nordlund 2002, Hennebelle & Chabrier 2008, Hopkins 2012). However, these models do not take into account the filamentary nature of core-forming clouds, and do not take into account the processes which disperse filament gas.

This paper explores the possibility that cores grow by gravitational contraction of their host filaments, and that the CMF arises from competition between this contraction and processes which limit the availability of filament gas for contraction. Filament gas can be dispersed by photoevaporation due to hot stars and by ablation due to shocks, winds and outflows. This competition between filament contraction and dispersal resembles the competition proposed to account for the IMF due to core contraction and dispersal (Basu & Jones 2004, Myers 2009, 2010), or due to protostar accretion and ejection (Bate & Bonnell 2005), in models of "stopped accretion" (Hennebelle & Chabrier 2011).

In this paper, Section 2 motivates a simple model of core formation, Section 3 presents the mathematical formation of the model, Section 4 gives results and application to nearby star-forming regions, Section 5 describes implications, Section 6 discusses limitations and compares with other models, and Section 7 gives a summary.

## 2. Core-forming filaments

This section describes properties of core-forming filaments, evidence for core growth by filament contraction, and limits on core growth due to filament dispersal.

### 2.1. Observed filament structure

Filamentary structure is a hallmark of observed star-forming clouds (Barnard 1907, Lynds 1962, Schneider & Elmegreen 1979), and the prevalence of filaments has become more clear with recent observations (Myers 2009, André et al 2010, Molinari et al 2010). Filaments develop in simulations of turbulent fragmentation, due to colliding supersonic shocks, with or without self-gravity (Klessen et al. 2004, Gong & Ostriker 2009, 2011). Filaments also arise from instabilities in self-gravitating sheets, depending on the strength and orientation of the magnetic field (Miyama et al. 1987).

Many observed filaments have complex structure with multiple components, branches, and cores. In simple cases their radial structure can be described by an isothermal or power-law model having a steep gradient, over a limited range of cylindrical radius. At larger radius the column density gradient is shallower and may approach a plateau configuration, as in the following examples.

In the filamentary gas associated with IC 5146, *Herschel* observations of dust emission from 27 filaments indicate that the column density $N$ typically declines steeply with projected radius $b$ from the filament axis, as for a self-gravitating isothermal cylinder, up to $b \approx 0.1$ pc. The column density then declines with shallower slope for radii extending out to more than 1 pc. For these filaments, the column density profile is well fit by the function

$$N = N_0[1 + (b/b_0)^2]^{-s} \qquad (1)$$

where $N_0$ is the peak column density, $s$ describes the steepness of the decline, and $b_0$ is a scale length at which the column density has declined from its peak by a factor $2^s$. The typical value of $s$ is 0.3, indicating a decline at large radius $N \sim b^{-0.6}$ (Arzoumanian et al. 2011). This decline is much shallower than in an isothermal model, where $s = 3/2$ and $N \sim b^{-3}$ at large radius (Ostriker 1964). This result corroborates a lower-resolution study based on near-infrared dust extinction (Lada et al. 1999).

This decrease of density and density gradient with increasing radius is also seen in other filamentary regions. In the Vela C region, *Herschel* scans across the brightest filamentary dust emission associated with RCW 36 indicate column density $N$ exceeding $10^{23}$ cm$^{-2}$ over a few 0.1 pc, declining as $b^{-0.8}$ to a plateau with $N = 5\text{-}8 \times 10^{21}$ cm$^{-2}$ (Hill et al 2011). Similarly, *Herschel* scans across filaments in the Pipe Nebula indicate column density profiles which are well fit by a Gaussian of typical width 0.08 pc, superposed on a plateau of a few $10^{21}$ cm$^{-2}$ (Peretto et al. 2012).

At smaller scales, high-resolution observations of $N_2H^+$ and $N_2D^+$ line emission from Oph N6 indicate a core which is elongated enough to be considered a small filament. Cuts of column density across the symmetry axis fit an isothermal cylinder model at 20 K on each side of the filament axis. Then the column density declines with a shallower slope on each side as the centrally condensed gas merges into the environment gas (Bourke et al. 2012).

The extended radial structure of the foregoing filaments, and of other observed filaments, implies that they have line density significantly greater than the criterion for isothermal equilibrium, and greater than that of the typical dense core (Inutsuka & Miyama 1992, 1997; André et al. 2010). These properties suggest that such filaments have more than enough mass to form cores, and that their self-gravity is sufficiently great to form cores through radial contraction.

**2.2. Filament contraction**

Cores form in contracting filaments according to numerical simulations by Bastien (1983), Rouleau & Bastien (1990), Inutsuka & Miyama (1997), Tilley & Pudritz (2003), Heitsch et al. (2009), and Gong & Ostriker (2011). The main formation mechanisms include gravitational contraction limited by magnetic fields and ambipolar diffusion (Tassis & Mouschovias 2007, Adams & Shu 2007), and dissipation of supersonic turbulence (Gong & Ostriker 2011). Cores may also grow by fragmentation of an equilibrium isothermal or polytropic filament (Larson 1985).

Core growth due to contraction of filament gas is also suggested by observations of infall asymmetry in nearby dense cores. In a recent study of 33 maps of the *CS J* = 2-1 emission line from starless cores in star-forming regions, 19 are dominated by spectra having blue-shifted distortion of the line profile, while three show evidence of oscillation and three indicate expansion (Lee & Myers 2011).

Most of these mapped regions (17 of 33) are parts of elongated filamentary clouds, and most of these maps (9 of 17) are dominated by infall asymmetry. In these maps, the infall asymmetry extends about 0.1 pc, or about twice as far as the core radius as seen in lines of $N_2H^+$ $J$ =1-0 (Lee et al. 2001, Lee & Myers 2011). Thus the inward motions extend beyond the core, according to most core definitions.

The mass accretion rate due to these inward motions is great enough to accrete the mass of a typical low-mass core in 0.5-1 Myr. Modelling of their spectral profiles indicates that the associated speed of inward motions has typical value 0.07 km s$^{-1}$ over the typical map radius 0.1 pc (Lee et al. 2001). For typical environment density $3 \times 10^3$ cm$^{-3}$ based on map size and extinction, the spherical mass accretion rate is $2 \times 10^{-6}$ $M_\odot$ yr$^{-1}$. This rate is essentially the same as that for a protostar forming from the collapse of a 10 K singular isothermal sphere, in the standard model of isolated low-mass star formation (Shu et al. 1987).

## 2.3. Dispersal of filament gas

Filaments in molecular clouds have a finite lifetime over which they can condense and provide mass to their cores. The lifetime of the typical filament in a cloud complex is at most a few Myr, since most of the stars in nearby cloud complexes are younger than ~4 Myr (Hartmann 2001) and since the oldest clusters with associated molecular gas are ~ 5 Myr old (Leisawitz et al. 1989).

The growth of a core due to gravitational contraction of its host filament can be halted if the surrounding gas becomes unavailable for significant accretion. This process is referred to here as "dispersal." To achieve such dispersal, it is not necessary to reduce the density of the surrounding gas to zero. Instead, it is sufficient to reduce the mean density of surrounding gas so that its free-fall time is several times the typical core lifetime. In that case the mass gain due to accretion of surrounding gas in the time available is nonzero, but is negligible compared to that already accreted.

Such dispersal can occur in several ways. The surrounding gas can be evaporated or ablated, so that the remaining environment gas is less dense. The surrounding gas can be shocked or heated, so that its velocity dispersion becomes comparable to the escape speed. The outward pressure gradient can increase due to magnetic forces or feedback from embedded stars. This section describes dispersal due to photoevaporation and ablation.

Progressive dispersal of filament gas is suggested by comparative studies of "globular filaments" (Schneider & Elmegreen 1979). These filaments can be viewed as an evolutionary sequence, extending from relatively young, undifferentiated cylinders to more evolved chains of isolated globules. This process is exemplified by the B213 filament in Taurus, which harbors five Class I and three Class II YSOs, having typical age less than 1 Myr. B213 has significantly less intercore gas than its neighbor filament B211, which has smoother structure and no associated YSOs (Schmalzl et al. 2010, Figure 12). Thus some of the initial filament mass of B213 may have dispersed while it formed stars, in the last ~ 1 Myr.

The best-studied mechanism of filament dispersal is photoevaporation due to dust heating by FUV photons from early-type stars, provided the local radiation field is sufficiently strong (Tielens & Hollenbach 1985, Kaufman et al. 1999, Gorti & Hollenbach 2002). Significant photoevaporation requires an increase over the mean interstellar radiation field (ISRF) by a factor $G_0$ of at least ~10, due for example to an early B star within a few pc. Then the outermost gas having column density ~ $10^{21}$ cm$^{-2}$ is heated and evaporates, exposing interior gas to further photoheating and evaporation.

Estimates of $G_0$ in nearby filamentary regions are ~1 in Taurus (Pineda et al. 2010), 3-5 in Lupus (Moreira & Yun 2002), ~ 4 in Corona Australis (Juvela et al. 2012), ~5 in IC 5146 within ~ 8 pc of the B0V star BD+46.3474 (Kramer et al. 2003), and ~400 in L1688 within ~ 0.4 pc of the B2V star HD147889 (Habart et al. 2003). These last two estimates imply $G_0 > $ ~10 within 6 pc of the B star in IC 5146 and within 3 pc of the B star in L1688.

The mass loss rate of a photoevaporating filament is of order $2\pi\Sigma\sigma l$, where $\Sigma$ is the mass column density of the heated gas, $\sigma$ is its sound speed, and $l$ is the filament length (Gorti & Hollenbach 2002). For $G_0 = 10$, gas with mean density 130 cm$^{-3}$ is heated to ~100 K (Kaufman et al 1999). Assuming these properties for a uniform cylinder segment having initial radial and axial extents 2.5 pc and 0.06 pc as in Section 5.1, the mass loss rate due to photoevaporation is 4 $M_\odot$ Myr$^{-1}$. This rate is comparable to the initial core mass gain rate due to contraction, as described in Section 3. At this rate it would take 0.5 Myr for evaporation to reduce the initial mass by a factor $e$.

The criterion $G_0 > $ ~10 for photoevaporation of a gas layer at ~ 100 K may be stricter than needed to limit the availability of gas for infall. More modest temperature increases in the outer layers of a filament can reduce the binding of the gas, even if they do not contribute to significant mass loss. Furthermore, such values of $G_0$ ~10 refer to nearby complexes forming relatively low-mass stars. Most stars form in clusters with more massive stars, and with much greater values of $G_0$ (Adams et al. 2006).

Filament gas can also be blown away (ablated), by shocks and winds which cause some of the filament gas to stream away in the flow direction. The time for a planar adiabatic shock to ablate a significant fraction of the mass of a centrally condensed cloud is a few times the cloud-crushing time scale $t_{cc} = \chi^{1/2} r_{c0}/v_s$, where $\chi$ is the ratio of the peak initial cloud density to the density of the intercloud medium, $r_{c0}$ is the scale length of the initial cloud density profile, and $v_s$ is the incident shock speed (Nakamura et al. 2006; see also Pittard et al. 2010). This time scale is estimated assuming the same ambient medium density as above, 130 cm$^{-3}$, and shock velocities 1-3 km s$^{-1}$ based on widths of CO lines. The initial filament model in Section 3 gives a scale length 0.04 pc at which the initial filament density has half its peak initial value of $2 \times 10^4$ cm$^{-3}$. For these values the ablation time scale is $t_{cc}$ = 0.2-0.5 Myr. Like photoevaporation, ablation tends to remove the outermost low-density gas, preventing it from accreting onto the core.

Thus photoevaporation and ablation have time scales 0.5-1 Myr for the filaments and environments considered here. The processes differ because photoevaporation is continuous once the filament is illuminated by sufficient FUV flux, while ablating flows can occur more impulsively, and can occur more than once in the lifetime of the filament. The stochastic arrival times of such flows may justify a probabilistic description of the stopping of core-forming accretion.

**2.4. Model assumptions**

The foregoing discussion justifies the basic assumptions of the model. Molecular cloud filaments are formed by colliding flows in a magnetized, turbulent medium and long-lived filaments are held together by self-gravity. As they become self-gravitating they develop the centrally condensed structure seen in observations and described above. As they differentiate into cores, they contract gravitationally in the radial and axial directions, adding mass to their cores. This process continues until the low-density gas around and between cores is accreted or dispersed. The mechanisms of dispersal include photoevaporation and ablation due to shocks and turbulent flows, with time scale 0.5 - 1 Myr. The competition between accretion and dispersal sets the rate and duration of the mass gain by each core.

**3. Filament structure and the CMF**

This section presents a simiple quantitative model following the above assumptions. The model first describes the CMF. This description and the distribution of accretion durations are then combined to obtain an estimate of the initial filament density profile. The procedure is similar to equating the mass of a protostar accreted as a function of time according to the IMF and ELS, to that due to gravitational contraction of a spherical cloud (Myers 2010).

This calculation proceeds in time-reverse order from a final state to an initial state, because the simple analytic form of the CMF allows a more convenient calculation of the density profile than in a time-forward calculation where a model of the density profile is used to generate the CMF.

This model neglects thermal, turbulent, and magnetic processes, so it should be considered at best illustrative of the competition between self-gravity and the processes which disperse dense gas. However, since the model is analytic it provides useful physical insight into a possible mode of core formation which is distinctly different from well-known models of turbulent fragmentation.

### 3.1. Mass function

Many observational studies of dense cores yield mass functions which resemble mass-shifted versions of the IMF, despite differences among these studies in resolution and in sensitivity to density and column density. It is therefore assumed here that the CMF and the IMF have the same basic form. A useful expression for this form is

$$\frac{dN}{d\log m} = \frac{\ln(10) N \Gamma \mu (\mu+\nu)^{-\Gamma}}{\nu} \qquad (2)$$

where $N$ is the number of objects, $m$ is the mass of an object, $\mu = m/m_0$ is the dimensionless mass normalized by the mass scale $m_0$, and $\nu = (1+\mu^2)^{1/2}$ is the quadrature sum of 1 and $\mu$. The exponent $\Gamma$ is the negative of the log-log slope of the high-mass tail of the distribution, first estimated for stars as $\Gamma = 1.35$ (Salpeter 1955, Bastian et al. 2010).

Equation (2) is a continuous approximation to the segmented IMF formulations of Kroupa (2002) and Chabrier (2005). With appropriate parameter values, it lies between the mass functions of Kroupa (2002) and Chabrier (2005), and it also lies between the continuous approximations to them given in Myers (2010). Equation (2) is similar to the expressions of De Marchi et al. (2010), Parravano et al. (2011), and Myers (2012), but it gives simpler expressions for the mass and mass accretion rate when combined with the ELS model for accretion duration.

The mass function in equation (2) provides a good fit to the most detailed CMF available, the distribution of 452 core masses in the Aquila complex (Könyves et al. 2010). Figure 1 shows the histogram of all Aquila cores with masses greater than 0.2 $M_\odot$, for which the completeness is estimated to be 75%. Cores with masses below 0.2 $M_\odot$ are less complete and are not shown here.

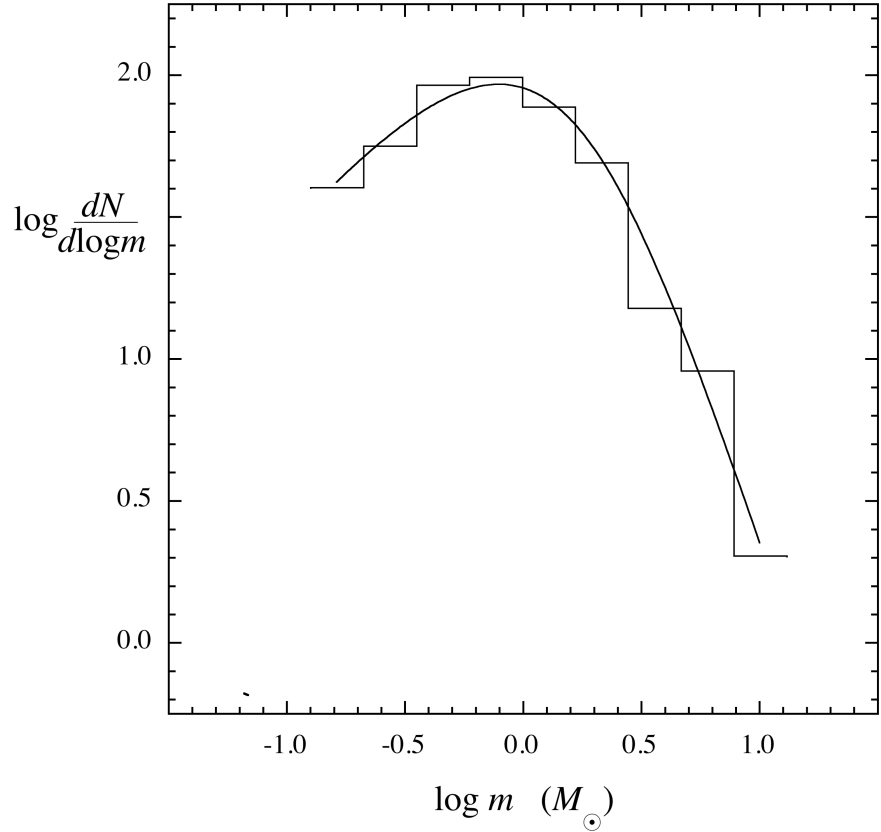

**Figure 1.** Starless core mass functions according to *Herschel Space Observatory* observations (*histogram*, Könyves et al. 2010) and according to the model equation (2) (*solid curve*) with mass scale $m_0 = 2.1\,M_\odot$ and exponent $\Gamma = 2.5$.

Figure 1 shows that the mass function in equation (2) matches the Aquila CMF well. Although this CMF is better determined than in other studies, it still allows significant variation among fit functions. The fit in Figure 1 has modal mass $0.79\,M_\odot$ while the modal mass for a log-normal fit is $0.6\,M_\odot$; the high-mass slope in Figure 1 has value $\Gamma = 2.5$ while the high-mass slope given by Könyves et al. (2010) is 1.5.

### 3.2. Mass and mass accretion rate

The model CMF in equation (2) and Figure 1 is equivalent to the probability density that the final mass of a core lies between $m$ and $m + dm$,

$$p(m) = \frac{\Gamma(\mu+\nu)^{-\Gamma}}{m_0 \nu} . \qquad (3)$$

The evolution of the core mass is described by assuming that a self-gravitating, centrally condensed "initial filament" contracts pressure-free from rest. This initial filament has much greater line density than the typical core, as is indicated by observations of filaments in IC 5146 (Arzoumanian et al. 2010) and in Serpens South (Kirk et al. 2012).

In this model a "core" is a coaxial segment of this contracting filament, defined by the adopted values of its mean density $\bar{n}_c$ and aspect ratio $\alpha$. As the filament contracts radially the radius $r_c$ which encloses mean density $\bar{n}_c$ increases. The length of the core is a fixed multiple $\alpha$ of this radius. Analysis of core maps indicates that $\alpha$ typically lies in the range 1-2 (Myers et al. 1991, Ryden 1996). The core mass $m_c = \pi \alpha m_p \bar{n}_c (r_c)^3$ also increases as the filament contracts. This process of core mass gain is referred to here as "accretion" even though it differs physically from the mass gain of a pointlike protostar from an extended medium.

As the filament contracts, its outer gas is subject to dispersal by evaporation or ablation as discussed in Section 2.3 above. Such episodes of evaporation and ablation can occur more than once and are expected to be uncorrelated in time. Thus it is useful to adopt a statistical description of the time when such filament dispersal stops core growth. For simplicity the stopping is assumed to be sudden compared to the typical growth time. The description adopted here is the same as is used for models of protostar accretion with "equally likely stopping" (ELS; Basu & Jones 2004, Bate & Bonnell 2005, Myers 2009, 2010).

In this ELS model the probability density that a core accretes from 0 to $t$ and stops accreting between $t$ and $t + dt$ is an exponential waiting-time distribution,

$$p(t) = \frac{\exp(-t/\tau_{stop})}{\tau_{stop}} \qquad (4)$$

where the accretion stopping time scale $\tau_{stop}$ is equal to the mean accretion duration in the limit where the accretion duration is much greater than the stopping time scale.

Many observed filaments have more than enough line density to form the typical dense core through radial contraction, and some filamentary clouds show evidence of contraction in the shapes of their spectral line profiles, as noted in Section 2.3. These properties suggest that some cores have mass limited more by the incomplete contraction of a relatively massive filament than by complete contraction of an relatively low-mass filament. If so, the final core mass depends more on the duration of the filament contraction than on the initial properties of the filament. This property resembles the finding that accretion age is more important in setting the mass of a protostar than differences in initial core properties, according to simulations of turbulent fragmentation (Bate 2009, 2012).

Assuming that accretion age is the most important factor in setting core mass, the probability densities of mass and accretion duration in equations (3) and (4) are related by

$$p(m)dm = p(t)dt \quad . \qquad (5)$$

Integration of equation (5), using equations (3) and (4), gives simple expressions for the mass as a function of time,

$$m = m_0 \sinh\left(\frac{t}{\tau_{acc}}\right) \quad , \qquad (6)$$

and for the mass accretion rate as a function of mass,

$$\dot{m} = \frac{m_0}{\tau_{acc}}\left(1 + \mu^2\right)^{1/2} \quad , \qquad (7)$$

where the accretion time scale $\tau_{acc}$ is related to the accretion stopping time scale $\tau_{stop}$ by

$$\Gamma = \frac{\tau_{acc}}{\tau_{stop}} \qquad (8)$$

as was also found in earlier IMF models based on ELS (Myers 2000, Basu & Jones 2004).

Equation (7) describes a mass accretion rate which approaches the constant value $m_0/\tau_{acc}$ for masses much less than the mass scale $m_0$. The mass accretion rate increases linearly with mass at large masses. This linear dependence of $\dot{m}$ on $m$ is an instance of "reduced Bondi accretion" in contrast to the dependence $\dot{m} \sim m^2$ in the standard treatment of spherical accretion onto a point mass (Bondi 1952, Zinnecker 1982).

The mass accretion rate in equation (7) is similar to the two-component accretion models in Myers & Fuller (1992), McKee & Tan (2003), McKee & Offner (2010), and Offner & McKee (2011). Following the nomenclature in McKee & Offner (2010) this mass accretion rate is referred to here as 2CA. This 2CA rate differs from these earlier rates because it is more explicitly related to the mass scale $m_0$ and slope $\Gamma$.

**3.3. Initial filament structure**

The initial filament structure is estimated by assuming that a core gains mass due to radial gravitational contraction of its parent filament, that a core stops gaining mass according to ELS, and that final core masses follow the CMF. The contraction is calculated in the pressure-free approximation. The difference between pressure-free and isothermal calculations is relatively small, as described in the Section 6.2.

Core growth is expected in both axial and radial directions, but it is expected that most of the mass gain of a core will come from radial contraction of the parent filament. Radial growth may be more important than axial growth because in nearby filamentary clouds, more filament mass appears to surround a core in the radial directions extending to the nearest filament, than in the axial direction extending to the nearest core. A recent study of flow along and onto the main filament in the Serpens South complex indicates that the radial mass accretion rate onto the filament is approximately three times greater than the axial rate along the filament (Kirk et al 2012). For simplicity the process of core mass gain is calculated here only for radial contraction.

The core extent in the axial direction is described by the parameter $\alpha$, but this extent is not modelled by a physical process. The axial structure is assumed to arise from axial nonuniformity in the initial filament, axial motions during contraction, and axially nonuniform dispersal of the filament gas.

3.3.1. Calculation of initial filament column density. The initial filament structure is calculated by assuming that the gas within cylindrical radius $r$ contracts radially from its initial value $a$, preserving its initial ratio $\lambda$ of mass per length. It is also assumed that the filament has initial central density equal to the defining core density $\bar{n}_c$, so that the core mass starts growing from zero mass at the start of filament contraction.

The initial mean density within radius $a$, $\bar{n}(a)$, is treated as a parameter which varies from 100 cm$^{-3}$ to the mean core density $\bar{n}_c$. Then the time $t_c$ is calculated for gas initially enclosed by radius $a$ to collapse until its mean density reaches $\bar{n}_c$, from the equations for pressure-free collapse of an infinitely extended cylinder (Penston 1969, Inutsuka & Miyama 1992, 1997). This time is equated to the time for the core mass to grow to $m_c$ according to the property that core masses subject to ELS follow the CMF, as in equation (6). Eliminating the core growth time gives the core mass, its initial radius $a$, and its initial density $n(a)$.

For comparison with observations of filament column density, the initial density $n(a)$ is integrated along the line of sight at offset $b$ from the filament axis, which is assumed to lie in the plane of the sky. After the line-of-sight coordinate is expressed in terms of $a$ and $b$, $N(b)$ is written as

$$N(b) = 2 \int_0^{b_{max}} da \frac{n(a)}{\sqrt{1-(b/a)^2}} \quad . \tag{9}$$

Here it is assumed that the maximum radial extent $a_{max}$ of the initial filament is the same as its maximum extent $b_{max}$ in the plane of the sky. In the comparison to observed filaments in Section 4, the value $b_{max}$ = 2.5 pc is adopted, matching the projected extent of IC 5146 filament 6 in Figure 2 of Arzoumanian et al. (2010).

## 4. Predicted and observed filament structure

The predicted structure of the initial core-forming filament is obtained according to the procedure described in Section (3) above, and from the choice of five associated parameters. The core mass scale $m_{0c}$ = 2.1 $M_\odot$ and the high-mass exponent $\Gamma$ = 2.5 are based on fitting equation (3) to the mass function of cores in Aquila (Könyves et al. 2010), as shown in Figure 1. The ratio $\alpha$ of core length to width is taken as 1.0 to give a nearly spherical core shape.

The two remaining parameters were adjusted to improve the match between the predicted and observed widths of filament column density profiles. The time scale for dispersal of core-forming filament gas is taken as $\tau_{stop}$ = 0.5 Myr, based on the discussion in Section 2.3. It was found that longer time scales extending to $\tau_{stop}$ = 1.0 Myr, gives initial filament widths which are narrower than typical values of 0.1 pc described by Arzoumanian et al (2010). Similarly, the mean core density is taken as $\bar{n}_c = 2 \times 10^4$ cm$^{-3}$, since greater values of mean core density, extending to $\bar{n}_c = 1 \times 10^5$ cm$^{-3}$ give initial filaments which are again narrower than typical.

These parameters $\tau_{stop}$ and $\bar{n}_c$ set the width of the initial filament because they set the ratio of its mean and peak density. The initial filament width whose enclosed gas has mean density equal to half its peak density is proportional to $\tau_{stop}^{-1/3}\bar{n}_c^{-1/2}$, based on the equations of core growth due to pressure-free cylinder collapse, core stopping due to ELS, and core masses following the CMF, as discussed in Section 3. If either $\tau_{stop}$ or $\bar{n}_c$ is increased too much, the resulting width decreases below the value needed to match observations.

With the foregoing parameter values chosen, the predicted profile is shown in linear and in logarithmic form in Figures 2 and 3, respectively. The profile is centrally condensed and narrow, with a peak value $N_0 = 8.0 \times 10^{21}$ cm$^{-2}$, FWHM 0.10 pc, and a slowly declining base with column density $\sim 1 \times 10^{21}$ cm$^{-2}$ as $b$ exceeds 1 pc. In logarithmic form the profile is flat for projected radii less than $\sim 0.01$ pc, and then declines approximately as $b^{-0.6}$ at large radii.

These predicted profile properties are similar to those of filaments recently observed with the *Herschel Space Observatory*. The column density profile which describes these filaments is given in equation (1), based on equation (1) of Arzoumanian et al. (2010). This "observed" profile closely matches the predicted profile, for parameters $N_0 = 8.3 \times 10^{21}$ cm$^{-2}$, $b_0 = 0.014$ pc, and $s = 0.28$. The predicted and observed profiles are shown in linear form in Figure 2 and in logarithmic form in Figure 3.

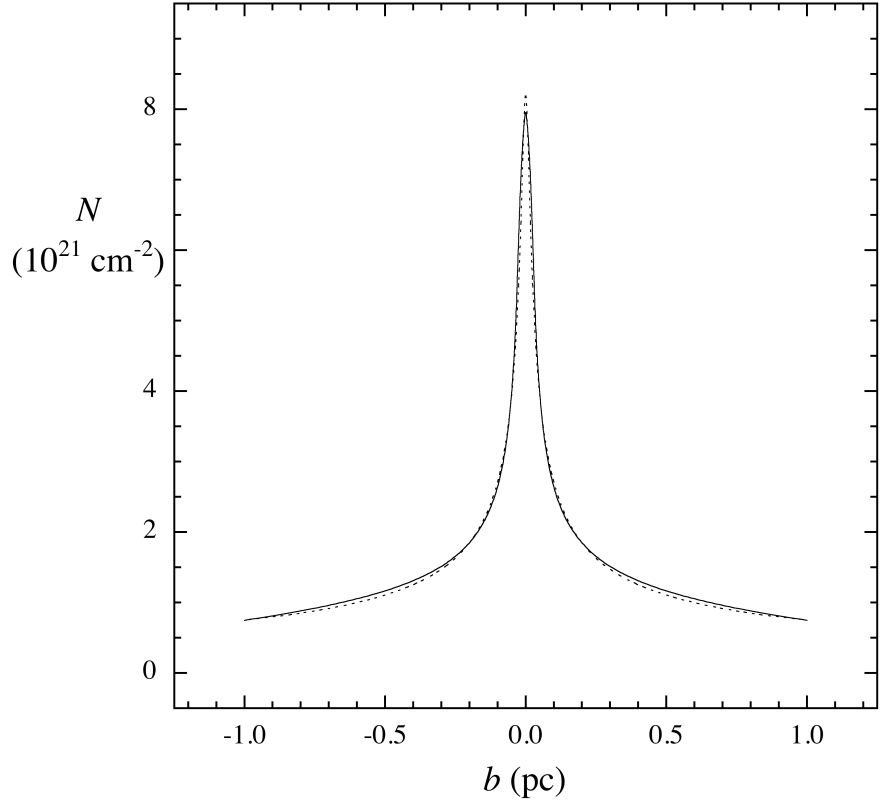

**Figure 2.** Predicted *(solid)* and observed *(dashed)* profiles of column density $N$ in a core-forming filament, as a function of projected distance $b$ normal to the symmetry axis. The predicted profile is the initial configuration of a filament which contracts by self-gravity until it is dispersed. The observed profile has the form which describes 27 filaments in IC5146 (Arzoumanian et al 2010). Each profile approximates that of an isothermal filament at small radii, with a shallower density decline at large radii.

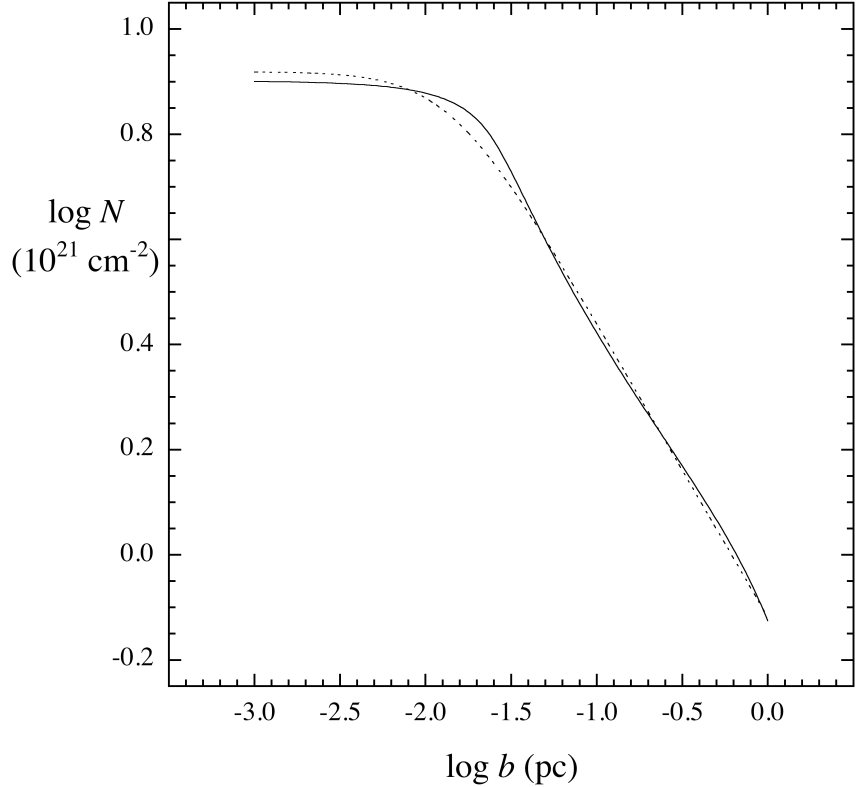

**Figure 3.** Predicted *(solid)* and observed *(dashed)* profiles of column density $N$ as a function of projected distance $b$, in logarithmic form, for the same predicted and observed profiles in Figure 3. Each profile has an inner region of nearly constant column density and an outer region of column density declining as $N \sim b^{-0.6}$.

The predicted and observed profiles are remarkably similar. They differ slightly in that the predicted initial filament has a larger central uniform zone than the observed filament profile. At 0.9 of the peak column density, the predicted width exceeds the observed width by a factor 1.6.

The best-fit observed column density profile in Figures 2 and 3 is also typical of the sample of 27 filaments in the IC 5146 complex, within statistical uncertainty. The FWHM filament widths $\Delta b = 0.1$ pc match exactly, and the best-fit column density scale length $b_0$, peak value $N_0$, and exponent $s$ each depart from the coresponding mean value in IC 5146 by -0.8, 1.6, and -0.3 sample standard deviations, respectively. The peak column density, $8 \times 10^{21}$ cm$^{-2}$, has the greatest departure from the typical value $3 \times 10^{21}$ cm$^{-2}$.

Despite these departures, the good agreement of the predicted and observed profiles in Figures 2 and 3, and the good agreement of the model and observed mass fucntions in Figure 1, suggests that the core formation model due to filament contraction and dispersal is consistent with observational constraints on the initial structure of core-forming filaments, and on the mass function of the cores formed.

The computation of the initial column density profiles in Figures 2 and 3 required a detailed analytic fit to the numerically calculated initial density profile. As a check, the initial density profile inferred from the CMF and ELS was used to predict the core mass as a function of time due to filament collapse. This was compared with the core mass as a function of time due directly to the CMF and ELS, in equation (6). The two curves agree within 3% for core masses from 0.09 to 10 $M_\odot$, verifying the calculation for the range of core masses under consideration.

## 5. Implications

### 5.1. Core properties

The core mass increases with time according to equation (6), until its mass gain stops due to filament dispersal. Then the core radius and mean column density grow with the same time dependence,

$$r_c = r_{0c} \sinh^{1/3}\left(\frac{t}{\tau_{acc}}\right) \tag{10}$$

and

$$\overline{N}_c = N_{0c} \sinh^{1/3}\left(\frac{t}{\tau_{acc}}\right) \tag{11}$$

where the core radius scale is

$$r_{0c} \equiv \left(\frac{m_{0c}}{\pi \alpha m_p \overline{n}_c}\right)^{1/3} \tag{12}$$

and the column density scale is

$$N_{0c} \equiv \frac{(\pi \overline{n}_c)^{2/3}}{2}\left(\frac{m_{0c}}{\alpha m_p}\right)^{1/3}. \tag{13}$$

The number distributions of core radius and of core mean column density are obtained from equation (2), since equations (11) and (12) indicate that the normalized radius and the normalized mean column density are simply related to the normalized mass $\mu$ by $r_c/r_{0c} = \overline{N}_c/N_{0c} = \mu^{1/3}$. The resulting distribution of core mean column density has modal value $\overline{N}_{c,\text{mod}} = N_{0c}\mu_{\text{mod}}^{1/3} = 6 \times 10^{21}$ cm$^{-2}$. The core radius distribution has the same shape as the mean column density distribution, and its modal value is obtained in similar fashion as $r_{c,\text{mod}} = 0.06$ pc.

These typical values of core radius and column density refer to cores defined by mean density $\overline{n}_c = 2 \times 10^4$ cm$^{-3}$. It is therefore useful to compare them to observed cores having a similar mean density. In a sample of 33 such starless cores the typical radius is $r = 0.07$ pc, based on observations of the (1,1) line of $NH_3$ (Jijina et al. 1999), and the corresponding mean column density is $(4/3)\overline{n}_c r = 6 \times 10^{21}$ cm$^{-2}$ for a spherical core.

Cores which have the same mass distribution as in Figure 1, but a mean density greater than $2 \times 10^4$ cm$^{-3}$ by a factor $g$ will have number distributions whose mean column density scale is increased by a factor $g^{2/3}$ and whose radius scale is decreased by a factor $g^{1/3}$, according to equations (12) and (13). Thus a core defined by mean density $10^5$ cm$^{-3}$ would have typical radius 0.04 pc and mean column density $2 \times 10^{22}$ cm$^{-2}$.

## 5.2. Age spread

The maximum accretion duration for the cores making up the CMF can be estimated from equation (8) for the core mass as a function of accretion duration. In the Aquila complex, the maximum core mass is ~10 $M_\odot$ (Könyves et al. 2010), indicating a maximum core accretion duration 2.8 Myr. This duration suggests that filaments in the Aquila complex first began contracting to form cores about 3 Myr ago. Then the oldest stars in this part of the Aquila complex should be younger than ~ 3 Myr.

On the other hand, the typical core accretion duration is ~ 0.5 Myr according to the accretion time scale discussed in Section 4. If some cores with such durations gave birth to protostars during their accretion, the youngest resulting YSOs should have ages of a few 0.1 Myr. This age spread of a few 0.1 Myr to a few Myr appears in many embedded clusters, which have numerous pre-main sequence stars and a smaller number of protostars (Myers 2012).

### 5.3. Infall speed

The model of core formation by filament contraction requires that the gas around forming cores have significant inward motions. The contraction speed at the core radius is obtained from the definition of the cylindrical mass accretion rate as

$$v_c = \frac{\dot{m}_c r_c \bar{n}_c}{2 m_c n(r_c)} \qquad (14)$$

where $n(r_c)$ is the density at the core radius. The density ratio $\bar{n}_c/n(r_c)$ is obtained from the density profile which yields the column density profiles in Figures 2 and 3. At the earliest times and smallest radii $\bar{n}_c/n(r_c) = 1$. For radii greater than the initial scale length, approximating the density profile as a power law gives $\bar{n}_c/n(r_c) = 3/2$. Following equation (14) the infall speeds at the core radius associated with core masses 0.2 - 10 $M_\odot$ are 0.05-0.08 km s$^{-1}$. These infall speeds are similar to those inferred from spectral line infall asymmetry in starless cores (Lee et al. 2001, De Vries & Myers 2005, Lee & Myers 2011).

It will be useful to carry out more detailed calculations of gravitationally contracting cylinders, including effects of thermal and magnetic pressure, to compare the expected line profiles with observations.

### 5.4. Filament structure and the CMF

The CMF in Figure 2 and the initial filament profile in Figures 3 and 4 are related according to the differing durations of core accretion described by the ELS distribution in equation (6). The lowest-mass cores with mass of a few 0.1 $M_\odot$ are the result of accretions of short duration, where only the innermost ~0.1 pc of the initial filament has time to accrete before dispersal. The cores of typical mass, ~ 1 $M_\odot$ near the peak of the CMF, are the result of accretion of typical duration, of order 0.5 Myr, from within initial radii ~0.2 pc. The very massive cores of several $M_\odot$ have accreted for the longest duration, of order 1-2 Myr, and they include gas of initial extent ~1 pc.

The most massive cores are of particular interest. In order to have such unusually long accretion durations, they probably began accreting earlier than most other cores, when dispersal due to stellar feedback was not yet significant. This head start for massive core accretion is similar to that suggested for massive protostars in clump-fed models of cluster formation (Smith et al. 2009, Wang et al. 2010).

Most of the gas incorporated into such massive cores originates in the shallow-slope wings of the density profiles in Figures 2 and 3. While low-mass cores arise from gas having the structure of an isothermal self-gravitating filament, cores in the high-mass tail of the CMF arise from initial gas with distinctly different structure. One may speculate that the innermost filament gas which makes low-mass cores has approached a thermalized, self-gravitating configuration, while the outermost gas which makes massive cores reflects the more dynamic conditions of filament formation and dispersal.

Therefore in this model, the shallow-slope wings of the initial density profile are crucial to the formation of massive cores. If instead the initial density profile were that of an isothermal self-gravitating cylinder, the steep profile slope would cause the CMF to have a finite maximum mass, whose value is proportional to the Jeans mass of the initial filament. This maximum mass is less than the critical mass for fragmentation of an isothermal filament (Larson 1985) by a factor $\pi^{3/2} 2^{-1/2} = 3.9$. This limit clearly contradicts the distribution of core masses in Aquila shown in Figure 1, since it would exclude a significant fraction of the cores in the distribution. Thus an isothermal initial filament could not make enough massive cores to be consistent with observations.

**5.5. Filament width**

As filaments contract radially, they can be expected to have progressively narrower column density profiles. If observations sample a range of contraction histories, one would expect to observe a corresponding distribution of profile widths. This expectation appears contrary to the finding that in nearby clouds, many observed filaments have similar column density profile width, 0.1 pc, in IC 5146 (Arzoumanian et al. 2010), in Vela and Serpens South (Hill et al. 2012), and in the Pipe Nebula (Peretto et al. 2012). To reconcile these two pictures it may be necessary to posit that most observed filaments have not contracted very much, or their internal structure is poorly resolved, or their distribution of widths is limited by internal thermal and magnetic pressure. It has been suggested that filaments may keep similar width if they maintain equilibrium as they slowly add mass (Inutsuka 2012).

It is also possible that filament widths appear similar in submillimeter wavelength images because externally heated filaments have relatively cold centers. *Herschel* estimates of filament dust temperature indicate decreases by a few K from outer to inner radii, e.g. from 15 K to 12 K (Arzoumanian et al 2010). This decrease is based on line-of-sight averages and is therefore probably a lower limit to the true decrease. If the true temperature decrease is sufficiently great, the submillimeter emission from central dust grains will be fainter than for a filament with no such temperature decrease. In that case, a submillimeter image will be biased against emission from the cold central part of the filament, and during contraction the apparent width may remain close to the initial width.

It may be useful to test this idea by comparing high-resolution column density profiles using both submillimeter emission, which depends on dust temperature structure, and near-infrared extinction, which is independent of dust temperature.

### 5.6. Modal mass

The CMF modal mass $m_{\rm mod} = 0.79\ M_\odot$ is obtained empirically by matching to the CMF in Figure 1, or to the initial filament density profile in Figures 3 and 4. A more physical explanation of the modal mass involves the competition between filament contraction and dispersal, which sets the core masses in this model. The modal mass depends on the initial density profile, which sets the available mass and the contraction time to reach the mean core density, for each initial radius. It further depends on the mean dispersal time, which selects the typical value of the largest initial radius which will survive dispersal.

This model therefore differs from fragmentation models because its mass scale depends on an initial condition, the initial density profile, and also on a final condition, the dispersal time scale, which specifies the typical duration of the core accretion.

It is notable that the modal line density $\lambda_{\rm mod} = \alpha\pi m_p n_0 r_{\rm mod}^2$ has the value 13 $M_\odot$ pc$^{-1}$ for the adopted parameter values. This line density is similar to the critical line density for fragmentation of an isothermal, self-gravitating cylinder at 10 K, 17 $M_\odot$ pc$^{-1}$ (Ostriker 1964). The densest filaments in the Aquila complex, which harbor cores and protostars, have line densities greater than this value, while other less dense filaments harbor fewer cores and protostars (André et al 2010).

### 6. Discussion

The model presented here gives a framework for understanding how observed CMFs can arise from filaments having observed radial density profiles, due to competition between gravitational contraction of the filament and dispersal of the outer filament gas. In this model, cores are the densest parts of their host filaments, and they are defined by their mean gas density and aspect ratio. Core size and mass increase as the filament contracts, until the supply of filament gas available for contraction is reduced by photoevaporation, ablation, and other mechanisms of "dispersal." For core mean density $2 \times 10^4$ cm$^{-3}$ the typical core size, mass, and mean column density are 0.06 pc, 0.8 $M_\odot$, and $6 \times 10^{21}$ cm$^{-2}$ respectively, and the typical infall speed at the core boundary is 0.05-0.08 km s$^{-1}$, each in reasonable accord with observed values.

The observed CMF is fit by a simple analytic function. When the accretion durations are described by equally likely stopping, the resulting core mass accretion rate is constant for low mass and varies linearly with mass for high mass. This two-component accretion rate is similar to those used to describe accreting protostars by Myers & Fuller (1992), McKee & Tan (2003), and Offner & McKee (2011).

The description of cores given here reflects their origin in filaments and allows the core mass to increase with time as the parent filament contracts. These properties differ from those of the static, isolated Bonnor-Ebert sphere, which is frequently used to model core structure.

The model results indicate a detailed connection between the structure of the initial filament and the shape of the resulting CMF. The high-mass slope of the CMF is closely related to the shallow slope of the filament density profile at large radii and low density.

## 6.1 Limitations

This model takes an observed filamentary profile as an initial condition, but does not explain how this profile arises. It remains for more detailed models of filament formation to account for the central condensation and the nonthermal wings indicated by recent observations.

It is oversimplified to assume, as in Section 3, that variations in structure from one filament to the next are negligibly small in their contribution of core mass, compared to the variation in accretion durations due to photoevaporation and ablation. A more realistic treatment should account for the observed range of filament properties, and not just their typical values.

It is also unclear whether the typical filament density profile should be taken as an initial condition at rest, as assumed here, or as an instance of filament contraction, as discussed by Arzoumanian et al. (2010). Contraction of an isothermal sphere with nonzero initial velocity has been discussed by Fatuzzo et al. (2004) and by Dalba & Stahler (2012). It may be possible to clarify the kinematic status of filaments with spectral line observations to infer radial and axial motions (Hacar & Tafalla 2011; Kirk et al. 2012).

The statistical model of equally likely stopping adopted in Section 3 matches the estimates of time scales ~ 0.5 Myr for photoevaporation and ablation given in Section 2, but details of filament gas dispersal are still poorly understood. New observations in the $^2P_{3/2}$ - $^2P_{1/2}$ line of [C II] may help to probe low-density, partly ionized gas associated with filaments (e.g. Mookerjea et al. 2012). It is important to understand how the lowest-mass cores arise in this picture, perhaps due to smaller initial filaments or more rapid dispersal processes.

The contracting filament model ignores many physical processes, and assumes an idealized initial structure, with no magnetic effects, no rotation, and no significant turbulence. It should be viewed as a tool to identify what physical processes are sufficient to explain the observations, and as a guide to more detailed investigations.

## 6.2 Pressure-free and isothermal collapse

The pressure-free model of filament contraction is sufficiently accurate for predicting the distribution of core masses. Pressure-free collapse is most accurate when the line density $\lambda$ of the infalling gas is much greater than the critical line density for an isothermal cylinder, $2\sigma^2/G$, where $\sigma$ is the velocity dispersion (Inutsuka & Miyama 1997). This condition is met for most of the gas in the present calculation, which has line density 44 $M_\odot$ pc$^{-1}$ within initial radius 1 pc, whereas the critical line density for 10 K gas is 17 $M_\odot$ pc$^{-1}$.

A more detailed comparison indicates that the mean gas density as a function of radius and time does not differ significantly between pressure-free and isothermal collapse, except for a relatively small range of radius and time. To compare these cases, an infinite cylinder was assumed to have the density structure of isothermal equilibrium. It then collapses according to the foregoing pressure-free model, and according to a self-similar isothermal model (Miyama et al. 1987, Inutsuka & Miyama 1992, 1997).

During the collapse, the pressure-free and self-similar solutions for the mean density are nearly the same if the initial radius $a$ is much greater than the thermal scale length $a_0$. Then each profile follows the scale-free profile of the singular isothermal sphere. Also, when the elapsed time approaches the central initial free-fall time, the mean density again tends to the scale-free profile, at all radii. Only for an intermediate time and at radii less than the initial scale length does the pressure-free mean density significantly exceed the isothermal mean density. At radius 0.03 pc, this excess is a factor $\sim 2$, or about one bin width in the typical log-log representation of mass functions.

**6.3. Comparison with other models**

Many models of the mass distribution of starless cores (CMF) have been made with the goal of matching the IMF; for a recent review, see Hennebelle & Chabrier (2011).

The best-known models of the CMF rely on gravoturbulent fragmentation, i.e. on properties of turbulence and on Jeans fragmentation. Padoan & Nordlund (2002) consider a superAlfvénic layer formed by turbulent ram pressure, where the number of cores formed by a velocity fluctuation at each scale $L$ is proportional to $L^{-3}$. They further assume a lognormal distribution of Jeans lengths to obtain a CMF similar to the IMF. Hennebelle & Chabrier (2008) assume a lognormal distribution of nonmagnetic density fluctuations, and select those satisfying a virial criterion to collapse and form prestellar bound cores. Turbulence provides nonthermal support which allows massive cores to survive without fragmentation. Hopkins (2012) extends this idea to associate the CMF with the mass spectrum of bound objects defined on the smallest scale on which they are self-gravitating.

In contrast to these CMF models of gravoturbulent fragmentation, the present CMF model is a model of "stopped accretion" (Hennebelle & Chabrier 2011). It has a more specific geometrical basis, since it relies on evidence that cores form in filaments (André et al 2010), and since it requires an initial density profile, which matches observations of filaments. It relies on an assumed distribution of accretion durations, based on estimates of gas dispersal timescales, whereas the gravoturbulent models do not describe how a growing fluctuation stops growing. It does not rely on properties of turbulent motions, such as the assumption that turbulent density fluctuations have a lognormal distribution. It does not assume a mechanism of fragmentation, although a fragmentation process may be necessary for a more complete explanation of how the filament develops into cores in the axial direction.

The relative simplicity of the present CMF model may be considered a drawback, since it ignores all but the most basic gravitational physics. On the other hand, the similarity of the shape of the CMF from cloud to cloud and from study to study suggests that its origin should be robust against most differences in cloud properties. In this model, the features which must be similar from region to region are the centrally condensed nature of the initial filaments and the distribution of their dispersal times.

## 6.4. CMF and IMF

The present model accounts for several observed features of filaments and cores, including the relation between filament structure and the CMF. It remains to better understand why the CMF and the IMF are similar, and how the cores in the present model generate protostars which also follow the IMF. These questions will be explored in a future paper.

## 7. Summary

This paper presents a model of core formation from filamentary molecular clouds. Its main features are:

1. A centrally condensed initial filamentary cloud matches detailed observations of filament radial structure, in the IC 5146 complex (Arzoumanian et al. 2010).

2. The initial filament contracts radially from rest due to its self-gravity.

3. A "core" is defined as a coaxial segment of the contracting filament having a specific mean density $\bar{n}_c$.

4. The core grows in radius and mass as the host filament contracts.

5. The filament contraction is calculated in the pressure-free approximation. Comparison with a similarity solution for isothermal contraction indicates that core masses are generally overestimated by a factor less than ~2.

6. The core stops growing when surrounding filament gas becomes unavailable for further accretion. Such dispersal is likely due to photoevaporation by FUV photons from nearby hot stars, and ablation by winds from nearby stars and by turbulent shocks.

7. Filament dispersal is described by a probabilistic model of equally likely stopping (ELS), whose time scale matches estimates for photoevaporation and ablation.

8. The competition between filament contraction and dispersal yields a distribution of core masses which matches detailed observations of cores in the Aquila star-forming complex (Könyves et al. 2010).

9. The shapes of the CMF and the initial filament density profile are closely related. The most massive cores derive most of their mass from gas in the nonthermal wings of the initial density profile.

10. The best match to the observed filament profile and the CMF arises when the mean core density is $\bar{n}_c = 2 \times 10^4$ cm$^{-3}$ and the filament dispersal time scale is 0.5 Myr.

11. The typical core size, mass, and mean column density are 0.06 pc, 0.8 $M_\odot$, and $6 \times 10^{21}$ cm$^{-2}$ respectively, and the typical infall speed at the core boundary is 0.05-0.08 km s$^{-1}$, each in reasonable accord with observed values.

12. The model initial filament has FWHM column density width ~ 0.1 pc as is typical of nearby complexes according to *Herschel* submillimeter observations. However, these same observations show fewer filaments with narrower widths than might be expected from filament contraction. Further high-resolution investigation of filament structure is needed, perhaps with observations of near-infrared extinction as well as with submillimeter emission.

**Acknowledgements**  I thank Fred Adams, Philippe André, Shantanu Basu, Tom Hartquist, Fabian Heitsch, Sun-Ichiro Inutsuka, Helen Kirk, Charlie Lada, and Ralph Pudritz for helpful discussions. The referee and Scientific Editor made suggestions which improved the paper. Irwin Shapiro and Terry Marshall provided encouragement and support.